\documentclass[times,twocolumn,final,authoryear]{elsarticle}

\usepackage{prletters}
\usepackage{framed,multirow}

\usepackage{amssymb}
\usepackage{latexsym}
\usepackage{amsmath}
\usepackage{algpseudocode}
\usepackage{subcaption}

\usepackage{natbib}

\usepackage{url}
\usepackage{xcolor}
\definecolor{newcolor}{rgb}{.8,.349,.1}

\usepackage{fancyhdr}
\begin{document}

\thispagestyle{empty}

\clearpage
\thispagestyle{empty}
\ifpreprint
  \vspace*{-1pc}
\fi

\begin{frontmatter}

\title{A Review of Cybersecurity Incidents in the Water Sector}

\author[1]{Amin Hassanzadeh}
\cortext[cor1]{Corresponding author}
\ead{amin.hassanzadeh@accenture.com}
\author[2]{Amin Rasekh}
\author[3]{Stefano Galelli}
\author[4]{Mohsen Aghashahi}
\author[5]{Riccardo Taormina}
\author[6]{Avi Ostfeld}
\author[7]{M. Katherine Banks}

\address[1]{Accenture  Labs, Cyber Fusion Center, 800 North Glebe Road, Arlington, VA.}

\address[2]{Zachry Department of Civil Engineering, Texas A\&M University, 400 Bizzell St, College Station, TX 77843.}
\address[3]{Pillar of Engineering Systems and Design, Singapore University of Technology and Design, 8 Somapah Rd., Singapore 487372, Singapore.}
\address[4]{Zachry Department of Civil Engineering, Texas A\&M University, 400 Bizzell St, College Station, TX 77843.}
\address[5]{Department of Water Management, Faculty of Civil Engineering and Geosciences, Delft University of Technology, Stevinweg 1, 2628 CN Delft, the Netherlands.}
\address[6]{Faculty of Civil and Environmental Engineering, Technion–Israel Institute of Technology, Haifa 32000, Israel.}
\address[7]{College of Engineering, Texas A\&M University, 400 Bizzell St, College Station, TX 77843.}

\begin{abstract}
This study presents a critical review of disclosed, documented, and malicious cybersecurity incidents in the water sector to inform safeguarding efforts against cybersecurity threats. The review is presented within a technical context of industrial control system architectures, attack-defense models, and security solutions. Fifteen incidents were selected and analyzed through a search strategy that included a variety of public information sources ranging from federal investigation reports to scientific papers. For each individual incident, the situation, response, remediation, and lessons learned were compiled and described. The findings of this review indicate an increase in the frequency, diversity, and complexity of cyberthreats to the water sector. Although the emergence of new threats, such as ransomware or cryptojacking, was found, a recurrence of similar vulnerabilities and threats, such as insider threats, was also evident, emphasizing the need for an adaptive, cooperative, and comprehensive approach to water cyberdefense.

\end{abstract}

\end{frontmatter}


\section{Introduction}

The water and wastewater sector (WWS) is considered by the US Department of Homeland Security (DHS) to be one of the main targets for cyberattacks among the 16 lifeline infrastructure sectors \citep{Whitehouse:2013}. Safeguarding it against cybersecurity threats is considered a matter of national priority \citep{Whitehouse:2017}. From 2012 to 2015, WWS received the highest number of assessments from the Cybersecurity and Infrastructure Security Agency Industrial Control Systems Cyber Emergency Response Team \citep{ICS-CERT:2016b}, which routinely conducts on-site cybersecurity assessments for several critical infrastructure sectors \citep{ICS-CERT:2016b}. The only exception was 2014, when the number of assessments in the energy sector was slightly higher \citep{ICS-CERT:2016b}.

According to ICS-CERT \citep{ICS-CERT:2016b}, 25 water utilities reported cybersecurity incidents in 2015, making WWS the third most targeted sector. Because there are over 151,000 public water systems in the United States \citep{epa2019information}, one may conclude that cybersecurity risk in WWS is low, and most systems are secure. However, the reality is that many cybersecurity incidents either go undetected, and consequently unreported \citep{walton2016}, or are not disclosed— because doing so may jeopardize the victim’s reputation, customers’ trust, and, consequently, revenues \citep{cava2018,rubin2019}. Moreover, the complexity and impact of cyber-originated incidents can be as serious as the incidents initiated from the operational technology (OT) area. Most industrial sectors, and WWS in particular, now are embracing the digital age, but still lack dedicated cybersecurity specialists to provide customized security guidelines, secure systems, and train employees.

Recently, cybersecurity has piqued the interest and attention of the WWS industry and policy-making entities. Several educational programs have been offered by the USEPA, DHS, the American Water Works Association, and the Water Information Sharing and Analysis Center over the last few years to raise awareness, train staff, and provide resources and tools to assist with cybersecurity practices \citep{waterisac:2015,US-CERT:2019,epa2019water}. This has been accompanied by a rising interest in the research community \citep{amin2013cyber,rasekh2016smart,ahmed2017model,formby2017out,taormina2017characterizing,laszka2017synergic,taormina2018battle,chandy2018cyberattack,taormina2018deep,housh2018model,ramotsoela2019attack}. In this respect, valuable lessons and insights may exist in past cybersecurity incidents that should be discovered and disseminated to inform the ongoing cyberdefense investments and efforts, thereby enhancing their relevance and effectiveness. This requires a comprehensive compilation and review of the these incidents, a public resource that is not currently available.

This study, conducted by the Environmental and Water Resources Institute (EWRI) Task Committee on Cyberphysical Security of Water Distribution Systems, presents a review of disclosed, documented, and malicious cybersecurity incidents in WWS to inform safeguarding efforts against cybersecurity threats. First, a review of a typical industrial control system (ICS) architecture, standard models, and common practices, alongside security controls and solutions offered for these environments, is provided. This is followed by a description of attack-defense models, which are an important concept in the design of cybersecurity systems. Next, a selection of cyber incidents in WWS is presented. The main details regarding the situation, response, remediation, and lessons learned are reported for each incident. This review concludes with recommendations for industry, policy makers, and research community.

\section{Industrial control networks}

In order to provide context for the analysis of the incidents, this section reviews traditional OT networks, their integration with Information Technology (IT) networks, and standard architecture designs proposed for ICS networks. We will refer to these architectures when reviewing some of the incidents and map the attacker's activities to the architectural layers and targeted hardware/software. 

\begin{figure*}[h]
  \centering
    \includegraphics[width=\textwidth]{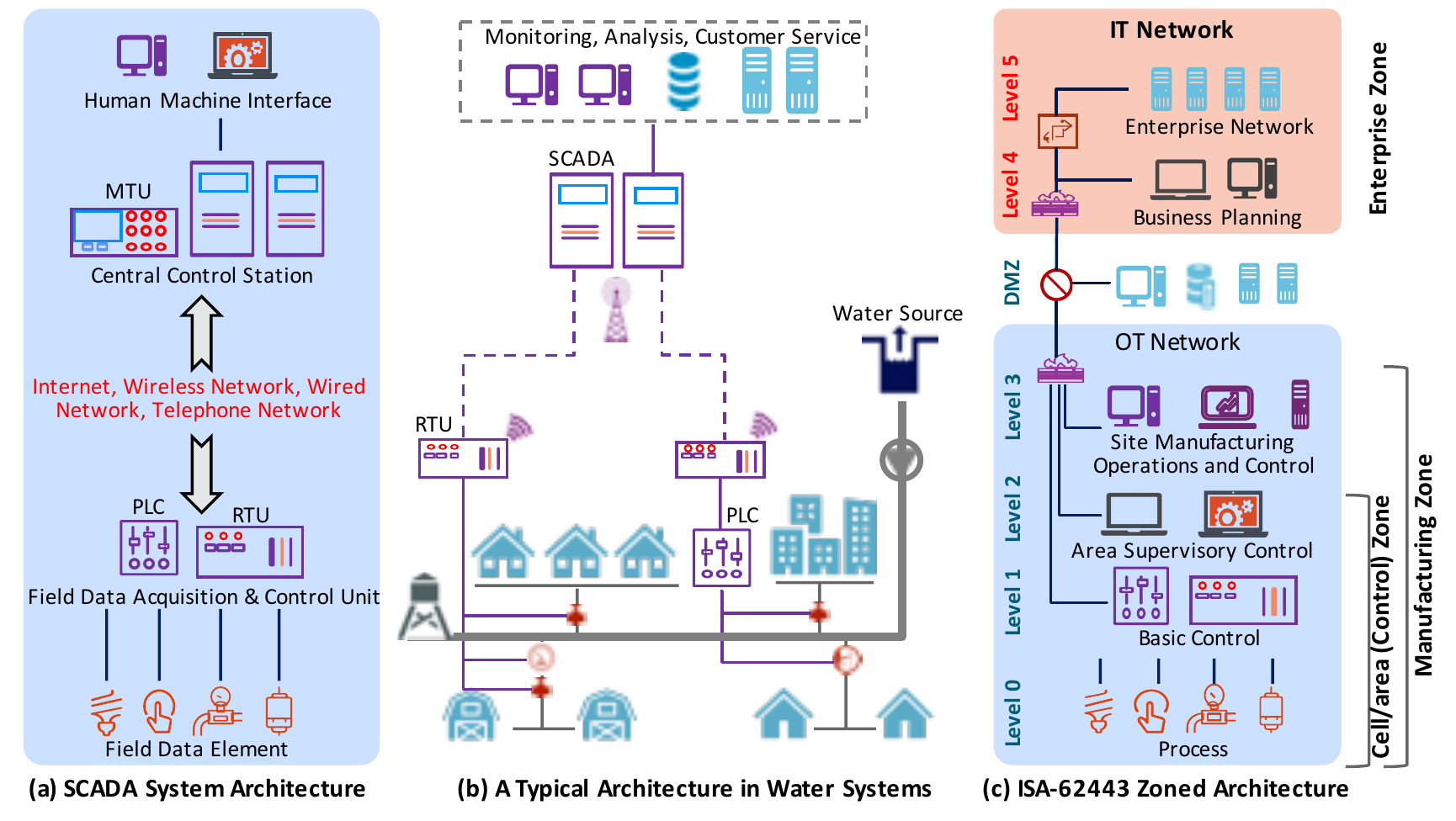}
  \caption[width=3cm, height=4cm]{(a) Traditional ICS architecture; (b) typical architecture in water systems; and (c) ISA-62443 zoned architecture.}
  \label{f1}
\end{figure*}

ICS networks traditionally uses a system of hardware and software components---called Supervisory Control and Data Acquisition (SCADA)---for process control, data collection, system monitoring, communication with industrial devices, and log data storing. A typical SCADA system architecture is depicted in Figure \ref{f1}a: the lowest level generally consists of field elements (also called end or dumb devices), such as sensors, pumps, and actuators. These elements are operated by control devices, such as Programmable Logic Controllers (PLC) and Remote Terminal Units (RTU). PLCs and RTUs are microcomputers that send control signals to the field elements, acquire data, and transmit them to the central control station, such as a Master Terminal Unit (MTU). MTU and RTUs/PLCs communicate and function in a master/slave model (through wired or wireless networks, public telephone network, or even through the internet) to send commands, upload new configurations, and monitor the field elements. Operators manage all these operations through a Human Machine Interface (HMI) connected to the MTU that allows them to gather data, send commands to remote sites, and change settings and configurations \citep{krutz2005securing, International:2009}. 

Figure \ref{f1}b shows a typical water system architecture with RTUs and PLCs geographically-dispersed in different sites. We have mapped different layers of a SCADA architecture to this sample network, where field elements, such as valves or pressure gauges, are monitored by RTUs with wireless antennas. The SCADA servers are located in a central control station (e.g., the headquarters of a water utility) and remotely communicate with the RTUs and PLCS scattered in the entire service area \citep{swan:2016}. 

For many years, SCADA systems, and, in general, OT networks in industrial environments, were air-gapped---that is, not connected to corporate IT networks or internet. However, as technology advanced, many organizations planned to consolidate overlapping IT and OT networks. This approach aims at saving maintenance costs and integrating data collection and analysis \citep{krutz2005securing}. However, such integration comes at high security risks due to the following reasons: 1) OT networks have different operational priorities compared to IT networks---e.g., availability vs. confidentiality---and one model may not fit both; 2) Most ICS devices and protocols are not designed to support security features like data encryption or access control, and often support remote access through radio modems; 3) Expensive legacy devices in ICS environments provide limited visualization options to implement and evaluate security modifications; and, 4) Critical and real-time business operations in OT, along with safety regulations, prevent immediate implementation of remediation options that may require system interruptions. In light of the above, security experts have proposed some work-around options to limit the access of users to the OT network. Other efforts in the ICS security field are constantly improving standards, protocols, and devices to support security features.

The new generation of converged IT-OT networks in industrial control systems, also referred to as Industrial Internet of Things (IIoT), is no longer air-gapped. Figure \ref{f1}c depicts a typical integrated ICS network consisting of multiple levels and zones, also known as the Industrial Automation and Control Systems (IACS) Security standard (ISA-62443) \citep{krutz2005securing}. A zone is in fact a set of assets (IT or OT devices) grouped together to provide a subclass of services and applications for the entire ICS network. The main zones can be described as follows:

\begin{itemize}
\item \textbf{Enterprise Zone} that includes assets for business logistics and enterprise systems, representing Level 4 and 5, respectively. This zone is also known as IT network. 
\item \textbf{Demilitarized Zone (DMZ)} that separates IT and OT networks, thus preventing direct access to OT devices from the IT network. All corporate–accessible services (e.g., web, email) reside in this zone.
\item \textbf{Manufacturing Zone and Control Zone}. The former refers to the entire OT domain, including Levels 0, 1, 2, and 3;  the latter refers to Levels 0, 1, and 2, so it is equivalent to the traditional ICS architecture shown in Figure \ref{f1}a. Level 3 provides site-level operation and asset management. Plant historian, production scheduling and reporting, patch and file services reside at Level 3 \citep{hassanzadeh2015towards}.
\end{itemize}

\section{Attack and Defense Models}\label{sec:attack_defense}

The incidents reviewed in this paper can be comprehended more effectively with some knowledge of attack and defense models, which are introduced next.

\subsection{Attack models}

From the attacker's perspective, a systematic process consisting of several steps or individual malicious activities is required to obtain the desired effect on the victim's network. Lockheed Martin researchers have expanded the kill chain concept used in military applications to define the Cyber Kill Chain (CKC) \citep{hutchins2011intelligence}, which models the life cycle of an attack based on the fact that the adversary uses a series of malicious activities (also called intrusions or single-step attack) and adjusts each step based on the success or failure of the previous step. CKC steps are defined as reconnaissance, weaponization, delivery, exploitation, installation, command and control (C2), and actions on objectives. Inspired by the CKC model, researchers have proposed several  attack life cycle models that are reviewed and discussed in \cite{hassanzadeh2015samiit}. 

\begin{figure*}
\begin{subfigure}{.5\textwidth}
  \centering
  \includegraphics[width=0.9\linewidth]{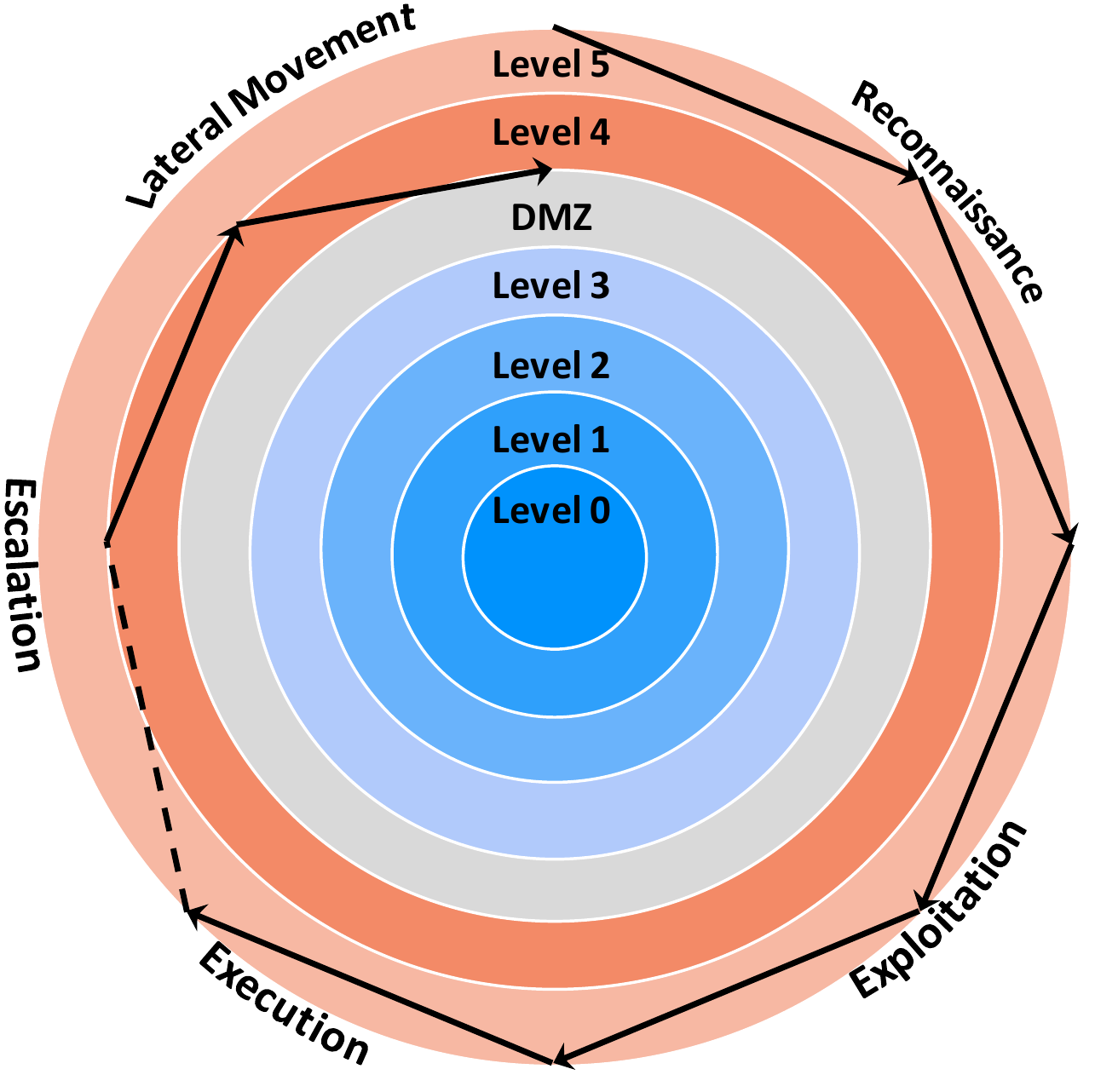}
  \caption{}
  \label{fig:samiit}
\end{subfigure}%
\begin{subfigure}{.5\textwidth}
  \centering
  \includegraphics[width=1.1\linewidth]{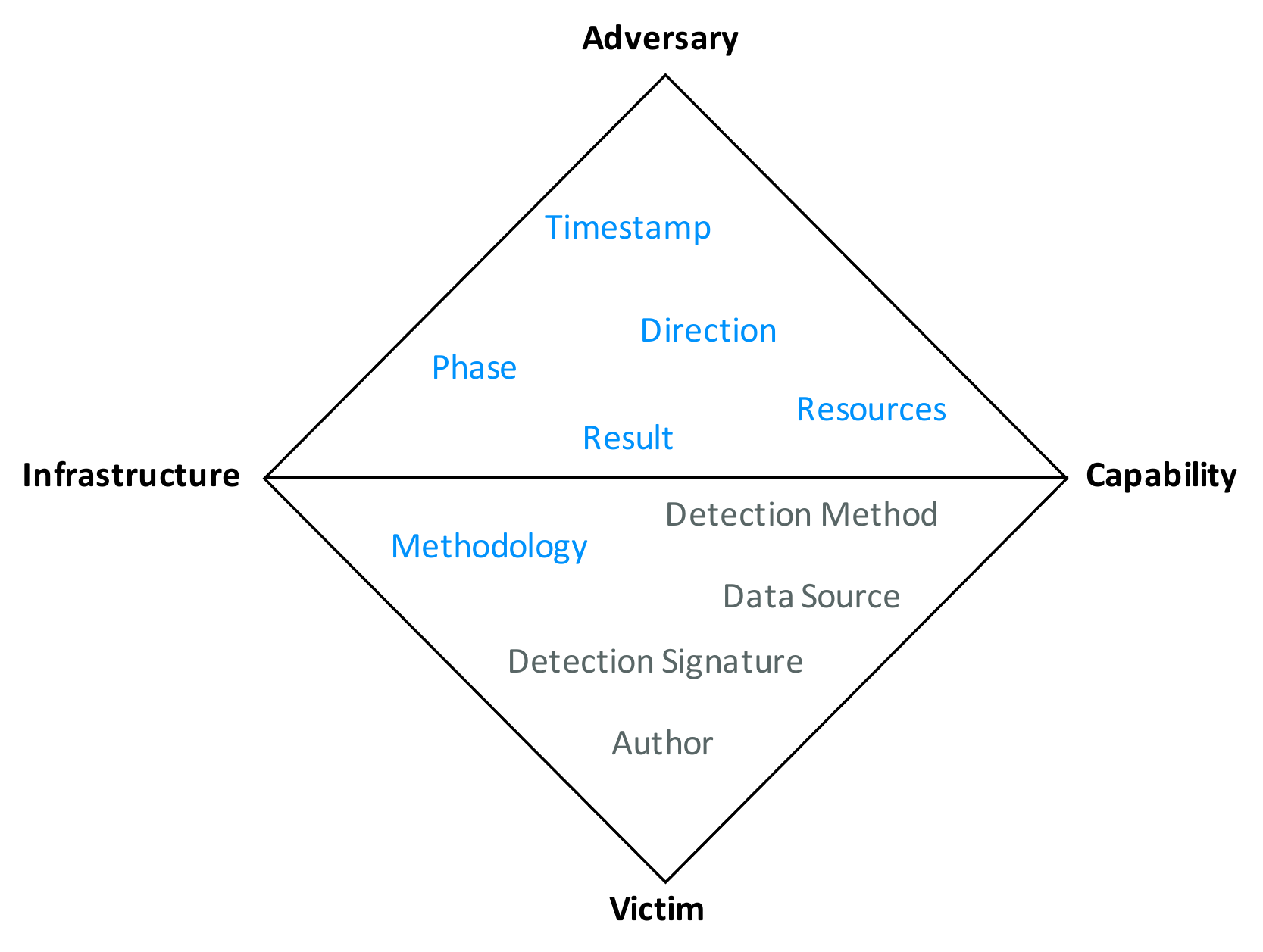}
  \caption{}
  \label{fig:diamond}
\end{subfigure}
\caption{(a) Spiral attack model in converged IT/OT networks; and (b) diamond model of intrusion analysis, with the core features at the corners, and metafeatures and expanded metafeatures inside the diamond.}
\label{fig:fig}
\end{figure*}

In industrial environments, the attack life cycle is slightly different because of the different architecture design shown in Figure \ref{f1}c. The target in such networks can be an asset in one of the three domains, namely, IT, DMZ, or OT. However, in most reported ICS incidents, the target is an OT asset \citep{hassanzadeh2015towards}, since the attacker gains access to the victim's environment through the IT domain and then traverse to the OT infrastructure by launching multiple attacks. This model is defined as the ICS Kill Chain, a multi-domain, multi-step approach that considers ISA-62443 architectural levels and CKC steps together. Since the attacker may need to repeat several CKC steps at each IT/OT level to laterally move within the network from one asset to another (until he/she reaches the target), \cite{hassanzadeh2015samiit} proposed a spiral attack model to accurately describe the attacker's activities within the converged IT/OT systems. Figure \ref{fig:samiit} shows a simplified version of this model, which is color-coded to map it to the IT/DMZ/OT domains of Figure \ref{f1}c. As depicted, an attacker may start with some reconnaissance activities in outer layers of an organization that are more exposed to the public (e.g., web server, mail server), and then find a vulnerable host that can be exploited. Once the first attack is delivered and executed, the attacker is already inside the victim's network, and then escalates his/her privileges and move laterally within the network towards the final target, which is placed in the lower levels. Note that this is a generic model, so there might be attacks that do not necessarily start from Level 5---such as an insider that uses OT workstations or a vulnerable server in the DMZ to launch an attack.

In light of the fact that an attacker operates in a chain of events (i.e., a set of single-step intrusions), the diamond model of intrusion analysis proposes a formal method called ``activity thread'' \citep{caltagirone2013diamond}. The method shows not only the attacker's steps and causal relation between them, but also a complete list of features for each of these steps. Figure \ref{fig:diamond} shows the core and meta features of each single-step intrusion, or event. An activity thread in an industrial environment is a directed graph (like the spiral set of arches in Figure \ref{fig:samiit}), where each vertex is an event/intrusion (see Figure \ref{fig:diamond}) and links represent the relation between those intrusions from the first step of the attack to the final target. As shown in Figure \ref{fig:diamond}, the four core features describe how an \textit{adversary} deploys a \textit{capability} over some \textit{infrastructure} against a \textit{victim}. Let us further focus on these features:

\begin{itemize}
\item \textbf{Adversary} is the actor or organization responsible for the attack. The adversary can be categorized as insider or outsider and individual, group, or organization. This is usually an unknown feature in most cyber-attacks. It is important to understand the distinction between adversary operator (i.e., the actual hacker) and adversary customer (i.e., the entity that benefits from the attack).

\item \textbf{Capability} is the set of tools and techniques that are used by the attacker. The vulnerabilities and configuration issues in the target environment define the capability of an attacker. 

\item \textbf{Infrastructure} is the physical and/or logical communication structure, such as email addresses or USB devices, used by the attacker to deliver the attack capabilities, maintain control over them, and finally obtain results. The infrastructure can be owned or controlled by the attacker or an intermediary (e.g., zombies hosts, botnets, or compromised email accounts).

\item \textbf{Victim} is the target that has vulnerabilities and configuration issues to provide attack capabilities for the adversary. Victims are either persona (e.g., people or organizations) or assets (e.g., networks, systems, accounts, or information).

\end{itemize}

In addition to the core features, there exist six meta-features in every security event: 1) \textit{timestamp}, that is, the start and stop time of the intrusion; 2) \textit{phase}, or \textit{step}, describing the position of the intrusion in the entire attack kill chain; 3) \textit{direction}, which denotes the course of an attack (for example, data exfiltration has a victim-to-infrastructure direction, while probing goes from the adversary to the infrastructure); 4) \textit{result}, which indicates the status of an attack, such as success, failure, or unknown; 5) \textit{resources}, such as software, hardware, information, knowledge, funds, etc.; and, 6) \textit{methodology}, that is, the class of the malicious activity, such as spear-phishing or denial-of-service. Moreover, four expanded-meta features have also been used to describe a single-step intrusion: \textit{detection method}, showing what tools or techniques were used in detecting the malicious activity; \textit{data source} to detect it; \textit{detection signature}, or \textit{rule}, that was used for the detection; and, \textit{author}, namely the analyst-author of the intrusion. Several multi-step attack examples and their activity threads are presented in \cite{caltagirone2013diamond}.

\subsection{Defense models}

To secure target organizations, defenders can employ several security tools and technologies. Moreover, they may have access to standards, threat intelligence databases, security controls, and benchmarks. Nonetheless, developing and implementing a thorough security strategy is a very challenging task that requires prioritization and rigour. The Center for Internet Security (CIS) proposed a list of the most fundamental and valuable security actions called ``CIS Controls'' that every organization should consider \cite{cis2019controls}. These controls are categorized as:

\begin{itemize}
\item \textbf{Basic Controls}, such as inventory and control of hardware/software assets, continuous vulnerability management, or controlled use of administrative privileges;

\item \textbf{Foundational Controls}, such as email and web browser protections, malware defenses, or secure configuration for network devices like firewalls, routers, and switches;

\item \textbf{Organizational Controls}, such as the implementation of a security awareness and training program, incident response and management, penetration tests, and red team exercises.

\end{itemize}

\begin{table*}[h]
\caption{List of CIS Controls}
\label{tb:cis}
\centering
\begin{tabular}{clc}
  \hline\hline
  \# & Security Control & Category\\ 
  \hline
  1 & Inventory of Authorized and Unauthorized Devices & Basic\\
  2 & Inventory of Authorized and Unauthorized Software & Basic\\ 
  3 & Secure Configurations for Hardware and Software on & Basic\\
  &Mobile Devices, Laptops, Workstations, and Servers&\\
  4 & Continuous Vulnerability Assessment and Remediation & Basic\\ 
  5 & Controlled Use of Administrative Privileges & Basic\\ 
  6 & Maintenance, Monitoring, and Analysis of Audit Logs & Basic\\ 
  7 & Email and Web Browser Protections & Foundational\\ 
  8 & Malware Defenses (installation, spread, and execution) & Foundational\\
  9 & Limitation and Control of Network Ports, Protocols, and Services & Foundational\\ 
  10 & Data Recovery Capability (information backup process)& Foundational\\
  11 & Secure Configurations for Network Devices such as Firewalls, & Foundational\\
  &Routers, and Switches&\\
  12 & Boundary Defense (detect, prevent, and correct unauthorized & Foundational\\
  &information flow) &\\
  13 & Data Protection (prevent exfiltration \& ensure integrity and privacy) & Foundational\\
  14 & Controlled Access Based on the Need to Know & Foundational\\
  15 & Wireless Access Control (track, control, prevent, and correct & Foundational\\
  & wireless accesses) & \\
  16 & Account Monitoring and Control & Foundational\\
  17 & Security Skills Assessment and Appropriate Training to Fill Gaps & Organizational\\
  18 & Application Software Security & Organizational\\
  19 & Incident Response and Management & Organizational\\
  20 & Penetration Tests and Red Team Exercises & Organizational\\
  \hline\hline
  \end{tabular}
\begin{tabular}{r}
\hline
\end{tabular}
\end{table*}

\begin{figure*}[h]
  \centering
    \includegraphics[width=\textwidth]{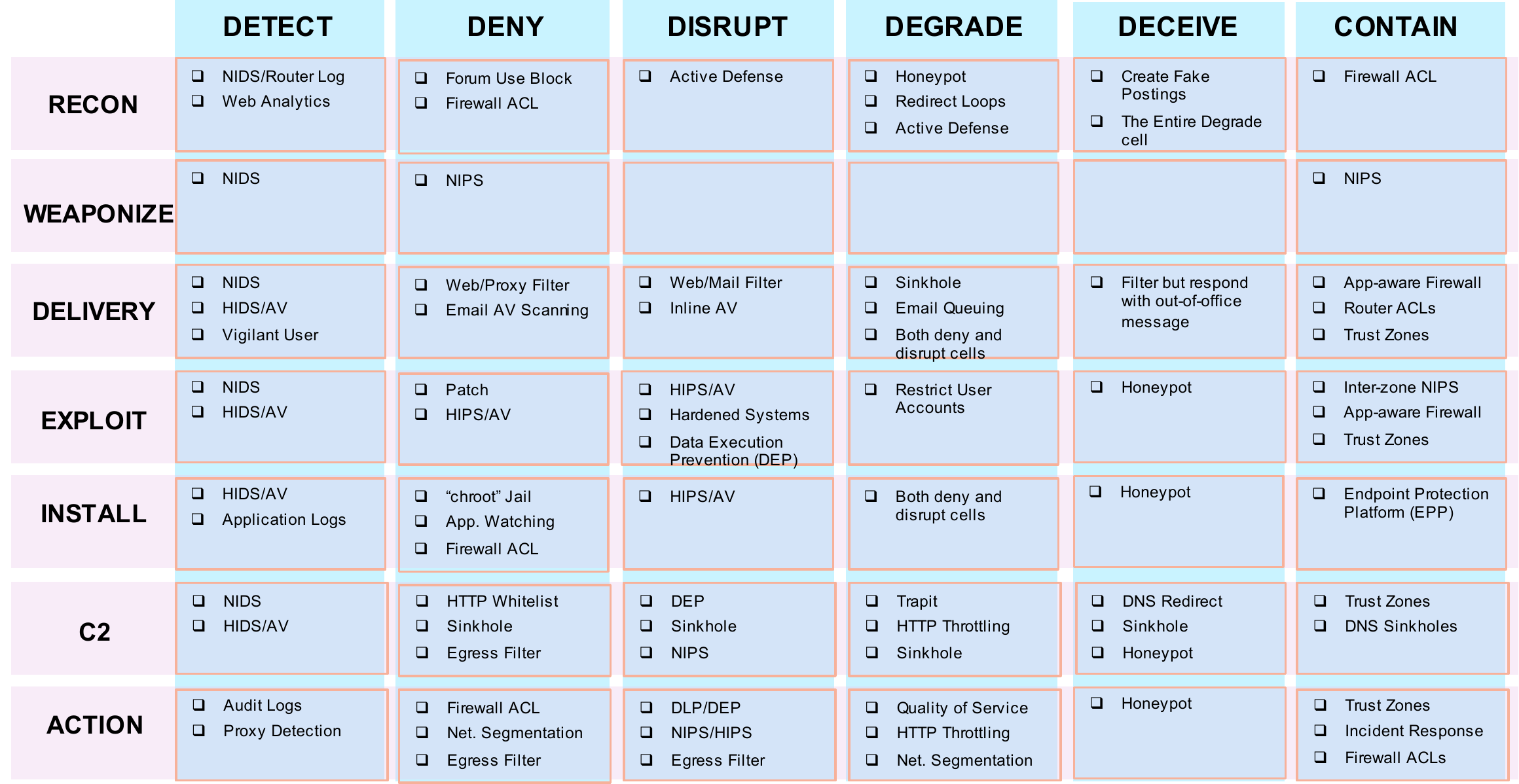}
  \caption[width=3cm, height=4cm]{Matrix of defensible actions at each step of an attack.}
  \label{CoA}
\end{figure*}

Table \ref{tb:cis} provides the complete list of CIS controls along with their corresponding category. These controls are available and offered in different security tools and solutions. They can have various impacts depending on their goal and implementation: 1) \textit{detect} the attack; 2) \textit{deny} or prevent the attacker from accessing assets or information; 3) \textit{disrupt} active malicious activities; 4) \textit{degrade} the impact of an attack; 5) \textit{deceive} the attacker; or, 6) \textit{contain} the malicious activity to a zone where damages can be mitigated. Figure \ref{CoA} shows how different security controls (tools and solutions) can be used to protect an organization against an intrusion attempt at each CKC step \citep{hutchins2011intelligence,bodeau2013characterizing,Nige2013DSP}. As an example, network-based intrusion detection systems (NIDS), host-based intrusion detection systems (HIDS), or anti-virus (AV) solutions can be used to detect exploitation activities. Similarly, trust zones can contain malicious activities associated with multiple attack steps from delivery to action, and honeypots can deceive attackers during several attack phases. AV solutions are mostly used to detect or disrupt attacks during the delivery, exploitation, or installation phase, while data execution protection (DEP) techniques are mostly used as a disruption mechanism.

In addition to traditional IT-based security controls, there exist several OT-specific security controls---such as data-diode and unidirectional gateway, in-line command white listing, passive asset discovery, passive OT intrusion detection (or anomaly detection), or patch and compliance management---that are currently used in industrial networks. A closer look at these solutions shows that they also fall under the categories mentioned above; however, they are designed to be compatible with OT network protocols and standards. For example, unidirectional gateway ensures a limited (if not zero) network interaction from the IT to the OT domain that should be considered as a firewall with a very restricted communication rule consistent with the OT architecture and its security needs. Hence, this OT-specific security control is a boundary defense control listed in Table \ref{tb:cis}. Similarly, passive asset discovery in OT networks is a basic security control to create an inventory of authorized and unauthorized devices (first control in Table \ref{tb:cis}). A technical report published by the Department of Energy \citep{DoE_21Steps:2005} lists 21 actions that can increase the security of SCADA networks. Each action corresponds to one or multiple security controls listed here.


\section{Incidents}

In this review, a cybersecurity incident refers to an incident that has been maliciously launched from the cyber space to cause adverse consequences to a target entity. All available reports on disclosed, documented, and malicious cybersecurity incidents in WWS happened until the end of May 2019 were considered, but only the incidents with detailed and verified information were then selected. The information sources include reports published by government organizations, scientific papers, internal reports from affected utilities, and media coverage that reported interviews with the involved official representatives. The authors of this review did not conduct any direct investigation themselves. The review is not restricted to any particular geographic region. All incidents, here presented in chronological order, are true positives, with the exception of one incident. This was included due to the massive, negative cry-wolf effects it created in the aftermath of its disclosure. For each incident, we describe the situation, response/recovery (if available), and lessons learned.

\subsection{Maroochy Water Services, Australia, 2000}

\subsubsection{Incident}

Maroochy Shire is located about 100 kilometres north of Brisbane in the Sunshine Coast region of Queensland, Australia. It has a population of nearly 120,000 inhabitants and a gravity sewage collection and treatment system that processes an average of 35 million liters of sewage each day. During the period 1997--2000, Hunter Watertech Pty Ltd (HWT), a third-party contractor, installed PDS Compact 500 RTUs at all 142 sewage pumping stations. This enabled to remotely control and monitor the pumps through a SCADA system. In late January 2000, the SCADA started experiencing faults, such as loss of communication and pump control capabilities, false alarms, or altered configuration of the pumping stations. The incident resulted in the release of nearly one million liters of raw sewage into the river, local parks, and residential grounds. 500 meters of open drain in a residential area were polluted. 

\subsubsection{Response and lessons learned}

In March 2000, after monitoring and recording all signals, the investigators concluded that the faults were caused by a human intervention. A suspect was caught on April 23$^{\text{rd}}$, 2000, having in his possession a Compact 500 computer, a two-way radio, a laptop, a transformer, and cables. The suspect had served as a site supervisor for Hunter Watertech until resigning due to unspecified disagreements (with effect from December 3$^{\text{rd}}$, 1999). He was sentenced to two years in jail and ordered to pay \$13,111 to the Council for the damage caused by the spill. The sewage spill and its impacts were cleaned up. The process took days and required the deployment of substantial resources. 

The main hazard involved in this incident was the unauthorized access to the SCADA system, which enabled the malevolent actor to release raw sewage into the surrounding environment. There were no cybersecurity procedures, policies, or defenses present, and the service contract was deficient or inadequate to handle the contractor’s responsibilities. Considering that the attacker was a former supervisor of the whole project, which controlled all pumping stations, the scale of the impacts could have been more extensive. The attacker was indeed a skillful, insider adversary with an intimate knowledge of the target system. The adoption of the NIST SP 800-53 control protocols \cite{bodeau2013cyber} would have arguably prevented all of the attacker's malicious activities. A former employee's access to the network, for example, should indeed be terminated immediately. [The sources used herein for this incident included \cite{COURT-QUEENSLAND:2002}, \cite{abrams2008malicious}, and \cite{sayfayn2017cybersafety}.]

\subsection{Pennsylvania Water Filtering Plant, U.S., 2006}

\subsubsection{Incident}

FBI suspected a security breach at a water treatment facility in Harrisburg, PA, in 2006. More specifically, it appeared that hackers planted a computer virus on the laptop computer of an employee. The hackers then used the infected laptop as an entry point, and installed a malicious software on the plant’s computer system. The hackers were reportedly operating outside the US. The investigations further reported that the hackers did not appear to target the actual plant, but merely intended to use the computer to distribute emails and other information. It was reported that the attack could have nevertheless affected the normal operations of the plant. For example, it could have altered the concentration levels of disinfectants in the potable water.

\subsubsection{Response and lessons learned}

The water utility eliminated remote access to the plant and changed all passwords. In the case of this specific attack, it should be noted that the entry point to the plant’s computer system was an employee’s laptop. Such weak links should always be avoided in the security chain. Due to the distributed nature of water infrastructure, staff often resorts to remote access to connect to key components and check system variables, such as tank water levels. Separating SCADA systems from administrative networks, which are connected to the internet, can decrease the risk of adversary penetrations. [The sources used herein for this incident included \cite{McMillan:2006}, \cite{epa2008water}, \cite{mcgurk:2008}, and \cite{risi:2019}.]

\subsection{Tehama-Colusa Canal, U.S., 2007}

\subsubsection{Incident}
The Tehama-Colusa Canal Authority (TCAA) consists of 17 water contractors of the Central Valley Project. Its service area spans across the west side of the Sacramento Valley. TCAA operated two canals in 2007---the Tehama Colusa Canal and the Corning Canal---that provide water for irrigation to a variety of permanent and annual crops in the local farms. Both canals are owned by the federal government. In 2007, a former electrical supervisor at the TCCA was alleged to have accessed and damaged the computer used to divert water from the Sacramento River to the local farms. Fortunately, the canals could still be operated manually. In his role with TCCA, the employee was responsible for the computer systems.

\subsubsection{Response and lessons learned}

The employee accessed the computer system around August 15$^{\text{th}}$, 2007, and installed unauthorized software on the SCADA system. He was an electrical supervisor with the authority and responsible for computer systems. The intrusion costed the TCAA more than \$5,000 in damages. The employee was eventually charged with unauthorized software installation and computer damage to divert water from the Sacramento River and sentenced to 10 years imprisonment and a fine. 

This incident is another case of insider attack. In this case, however, the insider was reportedly still an active employee of the affected entity at the time of the attack. [The sources used herein for this incident included \cite{McMillan:2007}, \cite{weiss2010protecting}, and \cite{risi:2019}.]

\subsection{Illinois Water Plant Pump Station, U.S., 2011 (a false alarm incident)}

\subsubsection{Incident}
In 2011, a pump burnout at an Illinois water plant was reported to be the result of a cyber-attack. News of the suspected attack became public after a security expert obtained a report collected by the Illinois Statewide Terrorism and Intelligence Center. According to the report, a plant’s  employee noticed problems in the SCADA. In particular, the pump kept turning on and off and eventually burnt out. The suspicions were raised in part due to the apparent connections to foreign IP addresses in the log files. This news was circulated rapidly by several credible news agencies. 

\subsubsection{Response and lessons learned}

The FBI and DHS launched an investigation. DHS spokesman subsequently advised that “At this time there is no credible corroborated data that indicates a risk to critical infrastructure entities or a threat to public safety”. According to the DHS, the pump had malfunctioned multiple times during the recent years. Additionally, the contractor with remote access to the computer system was on a personal trip in Russia. Investigation of the log files and interviews with the personnel collectively concluded that the reported attack was a false alarm.

Interestingly, this false alarm was circulated extensively by some credible news agencies, such as the Washington Post, causing anxiety and cry-wolf effects. The issue could have been prevented through a more timely consideration of the employee’s international travel and pump malfunctioning history. Another factor that likely contributed to the cry-wolf effect was the public availability of a preliminary report that anticipated the official conclusion of the investigations. [The sources used herein for this incident included \cite{nakashima2011} , \cite{zetter2011} , and \cite{parish011}.]

\subsection{Key Largo Wastewater Treatment District, U.S., 2012}

\subsubsection{Incident}

In 2012, the former Chief Financial Officer (CFO) of Florida’s Key Largo Wastewater Treatment District illegally accessed the district’s computer system to download emails and other personal documents. He performed these actions using the credentials of other employees, after the district did not renew his contract. He was arrested on account of felony charges, including computer crime with intent to defraud, modify information without authority, and delete information from the district’s computer system. 

\subsubsection{Response and lessons learned}
The facility’s IT manager discovered emails addressed to the CFO’s personal email account during a routine check of the email system. These emails were sent when the CFO was still working at the facility but already informed that his contract was not going to be renewed. Upon discovery, the IT manager informed the police, who then proceeded to arrest the CFO. The attack was limited to the IT systems of the facility, with no other malicious activity or disruptions for the district's operations.  

It is still not clear how the CFO got the credentials of his fellow employees. It is important for employees to constantly update their passwords in order to reduce the risks associated with stolen credentials. The CFO used these credentials to access the system from home, suggesting that no second authentication factor was needed to access the computer systems. Similarly to the `Kemuri Water Company' incident (Incident 8 below), a two-factor authentication could have prevented this attack. The attack was discovered thanks to routine checks, which should always been performed extensively for systems containing sensitive and confidential data. [The sources used herein for this incident included \cite{govtech:2012} and \cite{local10:2012}.]

\subsection{Bowman Avenue Dam, U.S., 2013}

\subsubsection{Incident}
The Bowman Avenue Dam is a small hydraulic infrastructure used to control floods in Blind Brook creek (Rye, New York). A key component of the dam is a remotely-controllable sluice gate, in operation since 2013, that controls the water flow as a function of water levels and temperatures in the creek. Between August 28$^{\text{th}}$ and September 18$^{\text{th}}$, 2013, hackers obtained “unauthorized remote access” to the SCADA system; a cyber-attack that allowed them to gather information on water levels, temperature, and the status of the sluice gate. The gate was manually disconnected for maintenance at the time of the intrusion, so hackers could not have the opportunity of taking direct control of the sluice gate. The attack was perpetrated with the aid of Google dorking, a computer hacking technique that leverages Google search engine to locate specific strings—and thereby vulnerabilities—in web applications, such as the one used to monitor and control the sluice gate. The hacker's action should not be classified as an intrusion, but rather as reconnaissance, namely the first stage of the CKC (see Figure~\ref{fig:fig}a), in which the attacker just gathers information on a potential target by looking for publicly available information on the Internet. The attacker used a standalone PC of the dam’s system to access its control network. However, at the time of attack, the control system was only gathering water level information and storing it on a spreadsheet. “The control system was attached to the Internet via a cellular modem but was directly Internet accessible and not protected by a firewall or authentication access controls.”.

\subsubsection{Response and lessons learned}

Since the attack, a new software and a new sluice gate have been installed. At Governor Cuomo's direction, New York State has taken multiple steps to improve its cybersecurity capabilities across several sectors. The investigations carried out by the DHS and Justice Department resulted in the indictment of a few state-sponsored hackers. The attack caused over \$30,000 in remediation costs. Whilst this attack had no consequences on the security and reliability of the Bowman Avenue Dam, it points to the vulnerabilities of critical water infrastructures, which are often monitored and controlled through unsafe web applications. It is thus not completely surprising to observe that the attack happened only two months after the intallation of an unsafe web application. [The sources used herein for this incident included \cite{nyc2016}, \cite{lach2016}, and \cite{kutner2016}.]

\subsection{Five water utilities, U.S., 2014}

\subsubsection{Incident}
In the spring of 2014, five water utilities across three states in the U.S. experienced some problems with their smart water meters. In particular, they faced inaccurate water bills and the deactivation of the Tower Gateway Base Stations (TGB), which receive signals from the water meters and transfer them to centralized facilities for monitoring and billing purposes. The first incident was reported by Kennebec Water District (Maine), where the utility could not connect to the TGB. Other nine attacks were reported in Spotswood (New Jersey), Egg Harbor (New Jersey), Aliquippa (Pennsylvania), and New Kensington (Pennsylvania).

The attack was caused by a fired employee of the company that manufactured the smart water meters---named company A in court's documents---who gained unauthorized access to protected computers. More specifically, the employee used to work as a field radio frequency engineer and was fired in November 2013. A few weeks later, using his access to the base station network, he conducted various malicious activities, such as changing the root passwords, modifying the TGB radio frequency, and overwriting computer scripts.

\subsubsection{Response and lessons learned}
 
 This abnormality drew the attention of the Federal government and caused investigations about possible cyber-attacks against the water infrastructures. Since the attack disabled the communication between utilities and their data collection network, the organizations had to resume manual data gathering. In addition, company A had to carry out forensic investigations at its own expenses to identify the attacker, characterize the attacks, and find and repair the damage.
 
 Though the utilities suspected that the disgruntled employee could have accessed the systems before May 2014, investigators could not link some anomalies to the attacker, since login details were not recorded at that time. However, recorded logins showed multiple intrusions linked to the IP address of the attacker’s home. The attacker was indicted for several malicious activities, and sentenced to prison and the payment of a fine.
 
 Even though the attacker was not a professional hacker, a default password allowed him to access the TGB. This highlights the importance of implementing access control and revoking access rights when someone is laid off. In addition, it is important to log and store in a safe place all logins and user’s activities. If company A had kept track of log-ins earlier, investigators could have discovered breaches dating prior to May 2014. This would have helped the investigations. [The sources used herein for this incident included \cite{doj:2017}, \cite{cimpanu:2017}, \cite{vaas:2017}, and \cite{Gallagher:2017}.]

\subsection{Kemuri Water Company (a pseudonym), U.S., 2016}

\subsubsection{Incident}

In 2016, an undisclosed water utility in the U.S. (presented under the pseudonym of Kemuri Water Company) hired Verizon Security Solutions to perform a proactive cybersecurity assessment of its water supply and metering system. A comprehensive assessment was subsequently conducted on both its OT (distribution, control, and metering) and IT (personal and billing information of the customers) systems. The assessment revealed several high-risk vulnerabilities, including a heavy reliance on outdated computers and operating systems. This included an outdated mid-range computer system (AS400) system that served a number of critical OT and IT functions---including the utility’s valve and flow control application---and had direct connections to many networks. 

The detection of these vulnerabilities triggered a full response and investigation. A cross-correlation of the utility’s internet traffic against a repository of known threat actors disclosed a positive match with the IP addresses of state-sponsored hacktivists. Interviews were also conducted with the utility’s staff: they revealed that some staff members have been aware of possible unauthorized access to the systems as well as a series of unexplained valve manipulation patterns. This casts doubt on whether the call for a forensic investigation was actually proactive and not reactive.  

A physical survey revealed the presence of a wired connection between the utility’s internet payment application and the AS400 system. Since the AS400 was open to the internet, it was concluded that access to the payment application would have also granted access to any information stored in the AS400. Collectively, the forensic investigations discovered an actual exploitation of the internet-facing payment application server and the subsequent manipulation of the utility’s valve and flow control application. In synthesis, the incident resulted in the exfiltration of 2.5 million unique records and manipulation of chemicals and flow rates. 

\subsubsection{Response and lessons learned}

Access to and from the account management web front was terminated, and outbound connectivity of the AS400 system was blocked immediately. Recommendations were made to replace the antiquated systems with more modern versions.

Multiple exploitable vulnerabilities led to the breach, which could have led to more serious consequences if the forensic investigation was not conducted earlier or the attackers had more knowledge of the utility's OT and IT systems. Internet-facing servers and applications, such the payment management application here, should not be connected to the SCADA. The utility had relied on a single-factor authentication; this is not sufficient, and multi-factor authentication should be used. Outdated systems, like the AS400 here, which formed a single point of failure, should not be deployed, and installation of security patches should not be overlooked. Exfiltration of records went unnoticed for a long time and in large amounts. There should be a monitoring mechanism in place that oversees the transfer of data to enable early detection and response. [The sources used herein for this incident included \cite{verizon:2016} and \cite{mahairas:2018}.]

\subsection{An undisclosed utility, U.S., 2016}

\subsubsection{Incident}
In 2016, the system administrator of a small water utility noticed the emergence of suspicious network traffic data. In particular, the administrator found heavy network traffic originating from the control panel of a pumping station. This triggered the possibility of a cyber-attack and a subsequent call to ICS-CERT. An official investigation was promptly launched.

\subsubsection{Response and lessons learned}
The ICS-CERT was immediately provided with the data on the network configuration. Address white-lists were instituted. Together with a transition to non-standard ports, these actions enabled safeguarding the network without requiring to put the control interface in offline mode. Within a few days, ICS-CERT also collected forensics images of the network hardware. Reverse engineering of the malware was subsequently performed to determine the attacker, breach point, data compromised, and mitigation strategy to prevent the same attack at other facilities. No details of the key findings have been disclosed. 

The situational awareness of the system administrator and prompt notification of ICS-CERT proved to be effective in isolating and thwarting a potentially catastrophic intrusion. Under the Critical Infrastructure Information Act of 2002 (CII Act), DHS has established the Protected Critical Infrastructure Information (PCII) Program to assure the utilities that their submitted information will not be disclosed. [The source used herein for this incident is \cite{ICS-CERT:2016a}.]

\subsection{An undisclosed drinking water utility, U.S., 2016}

\subsubsection{Incident}
In late 2016, an American water authority noticed a 15,000\% increase in their monthly cellular data bills. The authority was hacked between November 2016 and January 2017. The utility had seven Sixnet BT series cellular routers, which provided wireless access for monitoring the utility’s pumping stations as well as a few other sites. Four of these seven routers were compromised by the hackers. The hack was believed to be an opportunistic action to steal valuable internet bandwidth, resulting in the the authority’s cellular data bill soaring from an average of \$300 a month to \$45,000 in December 2016 and \$53,000 in January 2017. However, the intrusion did not damage the utility's infrastructure and did not cause any physical harm. The cause of the attack may stand in the Sixnet BT Series Hard-coded Credentials Vulnerability (identified by the DHS in May 2016). A poorly-skilled hacker should indeed be able of exploiting this vulnerability by hacking a factory-installed password. Sixnet produced patches and a new firmware to mitigate this vulnerability.

\subsubsection{Response and lessons learned}

The use of hard-coded credentials by the routers manufacturer and failure of the water authority to install the patches proved to be major contributors to this incident. [The sources used herein for this incident included \cite{walton:2017} and \cite{jerome:2017}.]

\subsection{A regional water supplier, U.K., 2017}

\subsubsection{Incident}
A regional water supplier was notified by several of its clients that their online account details were changed. After the clients credential were reset, it emerged that the details of some registered bank accounts were also changed, so that refunds issued to the customers were transferred fraudulently to these new bank accounts. In particular, the diverted refunds totaled over £500,000 and were directed to two bank accounts in England. The banks holding these accounts were socially engineered and allowed the holders to quickly transfer the majority of the funds to other bank accounts in Dubai and the Bahamas. Subsequently, these funds were used to purchase Bitcoins, which were then transferred to addresses associated with a Bitcoin mixing service, thus preventing any subject to be identified by following this trail further. 

\subsubsection{Response and lessons learned}
The company initially notified its legal advisor about the data breach. When the efforts to track down the bank account holders failed, the legal advisor contacted Verizon's cybersecurity experts, who started investigating in the company's premises. The experts proceeded to analyze the systems and processes involved in managing the customers' accounts. After a due diligence review of logs and web server revealed that no malicious software was present, the Verizon team suggested to interview personnel involved with customers' accounts. The interviews were extended to various stakeholders, including a third-party call center in Mumbai (India), which was responsible for administering the online accounts and processing telephone payments. After reviewing the Customer Relationship Management's log files, the investigators were able to confirm that one employee had accessed all the accounts that were fraudulently refunded. In depth analysis of the employee's computers revealed that, despite the use of a data wiping software, he had sent numerous email messages concerning the accounts affected by the fraudulent activity to another individual based in England. When presented with this evidence, the suspected worker finally confessed the crime and offered assistance in identifying accounts with over £1,000 in refunds stolen. The employee would take photographs of the account details and send them to his aide in England, who would then create an online account or request a password reset. With the help of the call center employee, new evidence was gathered, and authorities were able to secure a conviction also for the aide.

This insider attack examined here suggests that management should also ensure that partners having access to critical data perform stringent background checks on their employees. [The source used herein for this incident is \cite{verizon:2017}.]

\subsection{A European water utility, 2018}

\subsubsection{Incident}

A European water utility with a cloud-based OT analytics system hired a critical infrastructure security firm, Radiflow, to monitor its network. On January 21$^{\text{st}}$, 2018, suspicious network traffic was detected on the SCADA network. A series of new links to external IP addresses created a major network topology change, which triggered several alerts. The destination IP addresses were looked up, but this did not lead to any malicious site. Further investigation revealed that the addresses belonged to a “MinerCircle Monero Pool”. This led to the detection of crypto-mining malware in the OT network of the water utility. The investigation classified nearly 40\% of the traffic as related to mining operations, causing a 60\% surge in the overall bandwidth consumption. The investigation found no attempts of manipulating the controller configuration or sending commands.

\subsubsection{Response and lessons learned}
The security firm informed the water utility about the crypto-mining malware and infected servers. The recovery scheme included updating the anti-virus software on some servers as well as tightening the firewall security. The updated anti-virus software was successful in detecting the CoinMiner malware.

This incident is believed to be the first known instance of cryptojacking---i.e., the unauthorized use of a computing resource to illicitly mine cryptocurrency---being used against an ICS. Suspicious network traffic was the clue that led to the detection of the cryptojacking in this incident. Besides suspicious network traffic, high processor usage, sluggish response times, and overheating are some symptoms of cryptojacking that can be monitored for early detection. [The sources used herein for this incident included \cite{radiflow:2018}, \cite{newman:2018}, and \cite{kerner:2018}.]

\subsection{Onslow Water and Sewer Authority, U.S., 2018}

\subsubsection{Incident}
Onslow Water and Sewer Authority, a water utility company in Jacksonville (North Carolina) was targeted by cyber-criminals in October of 2018. Timed right in the wake of Hurricane Florence, the attack soon escalated into a sophisticated ransomware attack that locked out employees and encrypted databases, leaving the utility with limited computing capabilities. The hack began with persistent cyber-attacks through a virus known as EMOTET. With the EMOTET virus infection persisting, the authority reached out to outside security experts to investigate and respond to the attack.
At approximately 3 am on Saturday October 13$^{\text{th}}$, while the investigations were still underway, the malware launched a more sophisticated virus known as RYUK.  The IT team immediately disconnected the authority's facilities from the internet. Nevertheless, the situation soon exacerbated and the virus encrypted files and data.
The authority suspects that the attack has been a targeted one because the hackers chose a target that was recently hit by a natural disaster. Moreover, the sophisticated virus was launched at 3 am on a Saturday---a time in which the authority was most vulnerable. The authority soon received one email from the cyber criminals demanding payment to decrypt the damaged files and data. The authority dismissed the offer and stated it will not “negotiate with criminals nor bow to their demands.”

\subsubsection{Response and lessons learned}

The authority has been working with the FBI, the DHS, the state of North Carolina, and multiple security firms for remediation and recovery. The authority also planned to rebuild its IT systems from the ground up. 

The authority had multiple layers of protection in place, including firewalls and antivirus/malware software, when the hackers struck. Yet, their IT system proven to be penetrable. Ransomware is the fastest growing malware threat, targeting users of all types, according to the FBI. In this incident, the utility decided not to pay a ransom. This is in accordance with the federal guidelines---the US Government does not encourage paying a ransom to criminal actors. [The sources used herein for this incident included \cite{onwasa:2018} and \cite{mahairas:2018}.]

\subsection{Fort Collins Loveland Water District, U.S., 2019}

\subsubsection{Incident}
Fort Collins Loveland Water District serves customers in parts of Fort Collins, Loveland, Timnath, Windsor, and Larimer County (Colorado). On February 11\textsuperscript{th}, 2019, the staff of the Fort Collins Loveland Water District and South Fort Collins Sanitation District were unable to access technical data. Daily operations and customers' data were not believed to have been compromised. The utility had fallen victim to a ransomware cyber-attack. The hackers demanded a ransom to restore access (the amount of ransom payment demanded has not been disclosed to the public). The district declined to pay the ransom. 

\subsubsection{Response and lessons learned}
Within a few weeks, the district managed to unlock the data on its own. The decision on whether or not to notify the customers about the hack was also a challenge. Eventually, it was decided not to notify them, since the district did not store customers' data. All payments were indeed handled by a third-party vendor.

This is another case of ransomware attack in which the victim declined to pay a ransom. Data segmentation and segregation proven to be a helpful practice in safeguarding sensitive customer and daily operation data. Hiring a third-party vendor to handle customer payments prevented the customer data to be compromised. The practice of hiring third-party vendors, however, creates its own risks, as it was also manifested by Incident 11. [The sources used herein for this incident included \cite{ferrier2019} and \cite{sobczak:2019}.]

\subsection{Riviera Beach Water Utility, U.S., 2019}

\subsubsection{Incident}
On May 29\textsuperscript{th}, 2019, Riviera Beach, a small city of 35,000 inhabitants located north of West Palm Beach (Florida), was hit by a crippling ransomware attack after an employee of the police department opened an infected email. Paralyzing computer systems of the police department, city council and other local government offices, the ransomware sent all operations offline and encrypted their data. The attack also spread to the water utility, compromising the computer systems controlling pumping stations and water quality testing, as well as its payment operations.  

\subsubsection{Response and lessons learned}

A few days after the attack, the city council unanimously voted to authorize its insurer to pay 65 bitcoins, approximately \$600,000, to the attackers. The city would pay an additional \$25,000 as insurance deductibles out of its budget. Two weeks after the attack was disclosed, the IT department could bring the city's website and email services fully operational, while the water pump stations and water quality testing systems were only partially available. Although water quality sampling had to be performed manually, the city council’s spokeswoman assured that water quality itself was never in jeopardy. The FBI, Secret Service, and DHS investigated the attack and recommended the city not to pay the ransom. Regardless of paying the ransom, as of June 20\textsuperscript{th}, 2019, the sensitive data being encrypted by hackers were still inaccessible. 

While waiting for the attackers to share a decryption key, the local government authorized spending more than \$900,000 to buy new computer hardware---purchases which were planned for next year. According to a councilperson, most of the existing hardware was old and outdated, which made it vulnerable to the cyber-attack. In addition, the city’s computer network was not updated, and patches were not installed on time. 

It is known that local governments and small public utilities are less prepared for cyber-attacks, since they lack the budget and professionals needed to secure their IT and OT systems. That said, basic cybersecurity training raises awareness, and reduces the possibility of succumbing to devastating attacks unleashed by the naivety of uninformed employees, such as the case for Riviera Beach. Although paying a ransom looks like the easiest way to solve the problem, FBI and security experts suggest never to pay ransom as it only encourages future criminal activity. Preventing cyber-attacks from happening is always the best practice. [The sources used herein for this incident included \cite{Doris:2019}, \cite{Mazzei:2019}, and \cite{O'Donnell:2019}.]

\section{Discussion}
As outlined in the previous section, the complexity of cyber-incidents in WWS has increased during the last two decades. In some earlier incidents, such as the 2000 Maroochy Water Services hack, an insider simply and directly gained access to the OT controllers and performed malicious activities, while in some recent attacks, such the 2016 Kemuri Water Company hack, several IT and OT workstations were compromised by outsiders using multi-step attack techniques. In this section, we review and analyze some key points of the aforementioned incidents from both attacker and defender's perspectives.

Table \ref{tb:geo} provides an overview of the time, location, targeted systems type, the investigation teams (i.e., target organization, third-party security teams, or governmental agencies), and the impacts associated with each incident. The majority of targeted systems are US-based water systems, which might be because: 1) they use more advanced networking technologies (integrated IT/OT architecture) and are thus more exposed to the internet; 2) they are lucrative targets for hackers with a wide variety of goals; and 3) incidents reporting and information sharing is more systematically and extensively encouraged, required, and pursued in the US \citep{nist:2012}. There have been claims of WWS cyber-attacks in other countries, such as Ukraine \citep{martin:2018},  but limited reliable, information is publicly available for such incidents. The WWS systems targeted by the cyber-criminals have been very diverse, ranging from upstream water supply systems to downstream wastewater treatment plants, underlining the fact that all types of water systems are susceptible to cyber-attacks. Table \ref{tb:geo} also indicates that the consequences of the cyber-attacks have been extremely diverse. The attacks have led to the pollution of open water bodies, theft of irrigation water, data breach, and manipulation of chemicals rates in potable water, to name a few. No reports of human casualties was found by this study. It is also observed that the primary incident investigators rarely come from victim's organization. This might indicate a shortage of in-house security teams or trained personnel.

\begin{table*}[!htb]
\caption{Summary of the incidents.}
\label{tb:geo}
\centering
\begin{tabular}{rlrrrrr}
  \hline\hline
  \# & Location & Year & Target System & Investigator & Primary Impact\\ 
  \hline
  1 & Australia & 2000 & Wastewater & HWT \& Queensland EPA & Environmental pollution\\ 
  2 & PA, U.S. & 2006 & Water treatment & FBI & Data breach\\ 
  3 & CA, U.S. & 2007 & Irrigation & System personnel & Water theft\\ 
  4 & IL, U.S. & 2011 & Water plant & DHS & Cry-wolf effects\\ 
  5 & FL, U.S. & 2012 & Wastewater & System personnel & Data breach\\ 
  6 & NY, U.S. & 2013 & Dam & Justice Department & Data breach\\ 
  7 & U.S. & 2013 & Water utility & Third-party provider & Data manipulation\\ 
  8 & U.S. & 2016 & Water utility & Verizon Security & Control manipulation \\
  9 & U.S. & 2016 & Water utility &  DHS & Data breach\\
  10 & U.S. & 2016 & Water utility & DHS & Bandwidth theft\\
  11 & U.K. & 2017 & Water supplier & Verizon Security & Financial impact\\
  12 & Europe & 2018 & Water utility & Radiflow & Resource theft\\
  13 & NC, U.S. & 2018 & Water utility & State and Federal & Data loss\\
  14 & CO, U.S. & 2019 & Water district & System personnel & Denial of access\\
  15 & FL, U.S. & 2019 & Water utility & FBI, DHS and Secret Services & Data loss\\
  \hline\hline
  \end{tabular}
\begin{tabular}{r}
\hline
\end{tabular}
\end{table*}

Attackers are usually grouped based on their capabilities, motivations, and goals. Based on these characteristics, various groups of attackers are defined such as script kiddies (curious, unskilled individual), cyberterrorists (physical damage goals), cybercriminals (financial goals), hacktivists (social or political goals), and state-sponsored actors. It is worth mentioning that some other groups, such as cyber researchers, white/black hats and internal actors, have been also proposed in the literature \citep{ablon2018data}. Regardless of their goals and capabilities, attackers can be insider or outsider. Table \ref{tb:adversary} summarizes the type of attackers, their target assets and domains, and their final action on the observed target. Attacker and group for Incident 4 are not available simply because the incident was later confirmed to be a false alarm. It is observed that insiders are common adversaries in the water sector, as reported for the Key Largo Wastewater Treatment District, Maroochy Shire, Tehama Colusa Canal Authority, the five Eastern water utilities attacks, and a regional water supplier hack (Incidents 1, 3, 5, 7, and 11). This suggests that management and security teams should be more cognizant of changes in the behaviors of employees. For example, in the Maroochy Water attack, the attacker was no longer an employee. However, he still had access to the wireless network. Thus, he can be considered as an insider causing physical and financial damages (both cyber-criminal and cyber-terrorist) who changed the configuration of several OT controllers. In some similar examples, such as Incidents 3, 5, and 7, former employees or contractors tried to cause harm (financially or physically) through an unauthorized access to the IT or OT systems. In case of Incident 7, the attacker chose multiple targets in different domains of five utilities.

The attacker in the second incident was most likely a script kiddie (SK) outsider, who installed malware on the victim's computer to gain access to the internal information and distribute emails and information---there is no evidence of other groups of attackers in the public report. However, it is known that Attack 8 is performed by state-sponsored parties who targeted multiple IT and OT systems that resulted in the data exfiltration and manipulation of chemicals and flow rates. Incident 4 is known as a false alarm; however, several operational issues were observed at the same time, thereby confusing the investigation team. As shown in Table \ref{tb:adversary}, recent incidents (since 2017) appear to have a more complex nature. The attackers, insider or outsider, have been targeting databases, files, and account servers of the victims for financial purposes. As organizations advance and integrate their IT and OT systems and limit the OT systems from accessing to internet directly, the IT systems become of more interest for attackers and the entry point to the victim's network. The most interesting and unusual attack in this study is perhaps Incident 12, where attackers deployed a cryptocurrency mining code on the OT network of the target utility (most likely downloaded from malicious websites) to use the computational resources of OT machines as part of a mining pool that creates or discovers digital currency. 

\begin{table*}[!htb]
\caption{Adversary Analysis.}
\label{tb:adversary}
\centering
\begin{tabular}{rlrrrrr}
  \hline\hline
  \# & Attacker & Group & Target & Domain & Action\\ 
  \hline
  1 & Insider & C\&T & RTU/PLC & OT & Configuration Change\\ 
  2 & Outsider & SK & Workstations & IT & Data Exfiltration\\ 
  3 & Insider & C\&T & SCADA & OT & Software Installation\\ 
  4 & N/A & N/A & SCADA & OT & Physical process issue\\ 
  5 & Insider & Cybercriminal & Mail/File Server & IT & Data Exfiltration\\ 
  6 & Outsider & State-sponsored & SCADA/HMI & OT & Data Exfiltration\\ 
  7 & Insider & Cybercriminal & Multiple & IT and OT & Unauthorized Changes\\ 
  8 & Outsider & State-sponsored & Multiple & IT and OT & Multiple\\
  9 & Unknown & Unknown & SCADA & OT & Data Exfiltration\\
  10 & Unknown & SK & Routers & OT & Unauthorized access \\
  11 & Insider & Cybercriminal & Account DB & IT & Unauthorized access \\
  12 & Outsider & Cybercriminal & SCADA/HMI & OT & Cryptojacking\\
  13 & Outsider & Cybercriminal & Info. System & IT & Ransomware\\
  14 & Outsider & Cybercriminal & Databases & IT and OT & Ransomware\\
  15 & Outsider & Cybercriminal & Databases, SCADA & IT and OT & Ransomware\\
  \hline\hline
  \end{tabular}
\begin{tabular}{r}
\hline
\end{tabular}
\end{table*}

There is no single defense mechanism that can protect WWS against cyber threats, so the defense teams should use any mechanism (e.g., detect, deny, deceive) offered by critical security controls (CSC) \citep{cis2019controls} (see Table \ref{tb:cis}). In Table \ref{tb:defense}, we outline the most needed protection mechanisms and top-three basic and foundational CSC for the attacks described in this study. The foundational CSC are associated to specific architectural levels, based on the attacker's first step and weakest point of the victim's network. We note that in almost all incidents there exists a lack of organizational controls, such as ``Security Skills Assessment and Appropriate Training to Fill Gaps'' or ``Incident Response and Management.'' Although many organizations use proactive approaches---such as routine vulnerability and threat assessment or adversary simulation (red teaming - CSC 20)---to find security flaws in their network, most of the reviewed incidents were not detected proactively. Reactive security strategy, as seen in most industrial networks triggers, is ``respond when it happens.'' Table \ref{tb:defense} also shows that most of WWS networks suffer from a lack of preventive security mechanisms (column Deny in Figure \ref{CoA}), that is, the first line of defense in cybersecurity practice. 

\begin{table*}[!htb]
\caption{Defense Analysis.}
\label{tb:defense}
\centering
\begin{tabular}{rlrrrrr}
  \hline\hline
  \# & Approach & Protection & Basic CSC & Foundational CSC & Architectural Level\\ 
  \hline
  1 & Reactive & Deny & 1, 3, 5 & 12, 15, 16 & 1-2 \\ 
  2 & Reactive & Deny & 2, 3, 4 & 7, 8, 14 & 2 and 4\\ 
  3 & Reactive & Deny & 2, 3, 5 & 11, 14, 16 & 2-3\\ 
  4 & Reactive & Detect &  2, 5, 6 & 9, 11, 12 & 2-3\\ 
  5 & Proactive & Deny & 3, 5, 6 & 7, 13, 16 & 5 (or DMZ)\\ 
  6 & Unknown & Deny &  2, 4, 6 & 9, 11,12 & 2-3\\ 
  7 & Reactive & Deny & 1, 3, 5 & 14, 15, 16 & 2-4\\ 
  8 & Proactive & Detect & 1, 3, 4 & 9, 11, 14 & 2-5\\
  9 & Reactive & Disrupt & 2, 3, 4 & 8, 9, 13 & 2-3\\
  10 & Reactive & Deny & 3, 4, 5 & 11, 14, 15 & 3-5\\
  11 & Reactive & Degrade & 4, 5, 6 & 12, 13, 14 & 4-5\\
  12 & Proactive & Deny & 2, 3, 4 & 7, 8, 11 & 2-3\\
  13 & Reactive & Contain & 2, 3, 4 & 8, 10, 13 & 4-5\\
  14 & Reactive & Contain & 2, 3, 4 & 8, 10, 13 & 3-5\\
  15 & Reactive & Contain & 2, 3, 4 & 7, 8, 10 & 3-5 \\
  \hline\hline
  \end{tabular}
\begin{tabular}{r}
\hline
\end{tabular}
\end{table*}

\section{Epilogue}

This study presented a review of 15 cybersecurity incidents in the water and wastewater sector within the context of industrial network architectures and attack-defense models. The incidents covered a wide variety of vulnerabilities and situations. The incidents spanned over 18 years, from the Maroochy Shire Sewage Treatment Plant insider attack in 2000 to the Riviera Beach Water Utility ransomware attack in 2019. This review is an informative resource to guide securing of industrial control systems in WWS and other lifeline sectors against cyberthreats. The sheer diversity of the systems, attackers, and consequences associated with the incidents dictate a need for inclusive and comprehensive vulnerability assessments, as well as risk mitigation, preparedness, response, and recovery studies that account for such extreme heterogeneity.

Because the reports by official agencies indicated a large number of cybersecurity incidents in the WWS, this review may not be inclusive of all incidents. Many of them may not be made public. The framework developed by this study, however, was structured and designed so that it can readily accommodate extensions and updates as more incidents are possibly disclosed (or take place in the future). The development and maintenance of an online version of this repository is believed to be a significant future endeavor to pursue.

\section{Acknowledgments}
Mohsen Aghashahi and M. Katherine Banks are supported by Qatar National Research Fund (QNRF) under the Grant NPRP8-1292-2-548. Riccardo Taormina and Stefano Galelli are supported in part by the National Research Foundation (NRF) of Singapore under its National Cybersecurity R\&D Programme (Award No. NRF2014NCR-NCR001-40). Avi Ostfeld is supported by the EU H2020 STOP-IT project (Grant agreement ID: 740610).


\bibliographystyle{plainnat}
\bibliography{dfvae}


\end{document}